\documentclass[conference]{IEEEtran}
\IEEEoverridecommandlockouts
\usepackage{hyperref}
\usepackage{cite}
\usepackage{amsmath,amssymb,amsfonts}
\usepackage{algorithmic}
\usepackage{graphicx}
\usepackage{textcomp}
\usepackage{xcolor}
\usepackage{soul}
\usepackage{tikz}
\usepackage{pgfplots}
\usepackage[a4paper, total={184mm,239mm}]{geometry}
\def\BibTeX{{\rm B\kern-.05em{\sc i\kern-.025em b}\kern-.08em
    T\kern-.1667em\lower.7ex\hbox{E}\kern-.125emX}}

\newcommand*\circled[1]{\tikz[baseline=(char.base)]{
            \node[shape=circle,draw,inner sep=1pt,fill={rgb,255:red,233;green,113;blue,50},text=white] (char) {#1};}}
\renewcommand{\baselinestretch}{0.970}
\newcommand\copyrighttext{%
  \footnotesize \textcopyright 2026 IEEE. Personal use of this material is permitted.  Permission from IEEE must be obtained for all other uses, in any current or future media, including reprinting/republishing this material for advertising or promotional purposes, creating new collective works, for resale or redistribution to servers or lists, or reuse of any copyrighted component of this work in other works.
 
  Accepted for publications at Design, Automation and Test in Europe Conference (DATE) 2026.}
\newcommand{\copyrightnotice}{%
\begin{tikzpicture}[remember picture,overlay]
\node[anchor=south,yshift=10pt] at (current page.south) {\fbox{\parbox{\dimexpr\textwidth-\fboxsep-\fboxrule\relax}{\copyrighttext}}};
\end{tikzpicture}%
}
\begin{document}
\bstctlcite{IEEEexample:BSTcontrol}
\title{Late Breaking Results: CHESSY: Coupled Hybrid Emulation with SystemC-FPGA Synchronization
}
\author{\IEEEauthorblockN{
Lorenzo Ruotolo\IEEEauthorrefmark{1},
Giovanni Pollo\IEEEauthorrefmark{1}, 
Mohamed Amine Hamdi\IEEEauthorrefmark{1}, 
Matteo Risso\IEEEauthorrefmark{1},
Yukai Chen\IEEEauthorrefmark{2}, 
Enrico Macii\IEEEauthorrefmark{1}, \\
Massimo Poncino\IEEEauthorrefmark{1}, 
Sara Vinco\IEEEauthorrefmark{1},
Alessio Burrello\IEEEauthorrefmark{1}, 
Daniele Jahier Pagliari\IEEEauthorrefmark{1} 
}
\vspace{-0.3cm}
\IEEEauthorblockA{\\\IEEEauthorrefmark{1}Dept. DAUIN, Politecnico di Torino, Italy, name.surname@polito.it\
    }
    \vspace{-0.4cm}
\IEEEauthorblockA{\\\IEEEauthorrefmark{2} IMEC, Leuven, Belgium\
    }
    \vspace{-1cm}
}

\maketitle
\copyrightnotice
\begin{abstract}
The growing complexity of cyber-physical systems (CPSs) calls for early prototyping tools that combine accuracy, speed, and usability. 
Virtual Platforms (VPs) provide fast functional simulation, but hybrid co-emulation solutions, in which key digital components are deployed on FPGA, become necessary when accurate timing modelling is required and RTL simulation is too costly. %
However, existing hybrid emulation tools are mostly proprietary, and rely on vendor-specific FPGA features.
To address this gap, we introduce an open-source framework that connects SystemC-based VPs with FPGA emulation, enabling full-system co-emulation of digital and non-digital components. The FPGA accelerates the execution of main digital subsystems, while a wrapper coordinates timing and communication with the VP through JTAG, maintaining synchronization with simulated peripherals.
Evaluations using a RISC-V SoC, with an example in the biosignals processing domain, show up to 2500$\times$ speedup compared to RTL simulation, while maintaining less than 2$\times$ total simulation time relative to pure FPGA emulation.
\end{abstract}

\vspace{-0.25cm}
\section{Introduction and Background}
Early prototyping helps shortening time-to-market and reducing costs of Cyber-Physical Systems (CPS) design~\cite{cps-survey}. A key challenge lies in accurately modeling interactions between digital systems and their surrounding environment (e.g., sensors, actuators, power sources) within a unified framework. 
Standard approaches rely on high-level SW Virtual Platforms (VPs) such as QEMU or Renode, which enable fast, functional
simulation and facilitate multi-domain full-system modeling through frameworks like SystemC AMS~\cite{6176558,qemu, qemu_paper, renode, renode_paper, systemc, systemcams}. 
However, pure SW VPs trade timing fidelity for speed and require separately maintained high-level abstractions of each component. 

Conversely, FPGA emulation executes the actual RTL code of digital blocks with cycle accuracy, without incurring the huge costs of RTL simulation. However, it complicates the modelling of 
interactions with components not available at RTL, or non-digital~\cite{fpga-survey}. 
VP-FPGA hybrid co-emulation frameworks~\cite{scemi} address this duality by combining a SW VP with FPGA-accelerated modelling of key digital blocks (e.g., processors), but existing implementations either are proprietary/not openly available~\cite{seok2017hla, aldec,1193229, jungersythil, 4408260}, rely on vendor-specific features~\cite{ou2005matlab, chimera} or require RTL instrumentation (e.g., transactors), as for SCE-MI~\cite{scemi} and others~\cite{firesim, chimera, tan2022emunoc, jungersythil}.

To address these limitations, we propose \textbf{CHESSY} (Coupled Hybrid Emulation via SystemC–FPGA Synchronization), a \textit{fully open-source}, flexible, hybrid prototyping framework that combines FPGA-accelerated RTL execution with loosely timed SystemC models~\cite{messy}. CHESSY leverages JTAG for VP-FPGA communication and timing synchronization, through lightweight Board Support Package (BSP) stubs, thus being \textit{FPGA-agnostic and requiring no changes to firmware or RTL}\cite{1193229}. The result is a portable and transparent co-emulation environment that bridges cycle-accurate RTL execution of key digital modules with high-level simulation of the rest of the system, including non-digital components.

We demonstrate CHESSY on a RISC-V System-on-Chip \cite{astral, pulp} for a biosignal-based robot control use case. The results show that we achieve more than three orders of magnitude speedup over RTL simulation while maintaining a total simulation time of less than 2$\times$ that of pure FPGA emulation.
CHESSY is available at \url{https://github.com/eml-eda/chessy}.

\vspace{-0.2cm}
\section{Methodology}
\label{sec:methodology}
\begin{figure}[t]
    \centering
    \vspace{-0.1cm}
    \includegraphics[width=\linewidth]{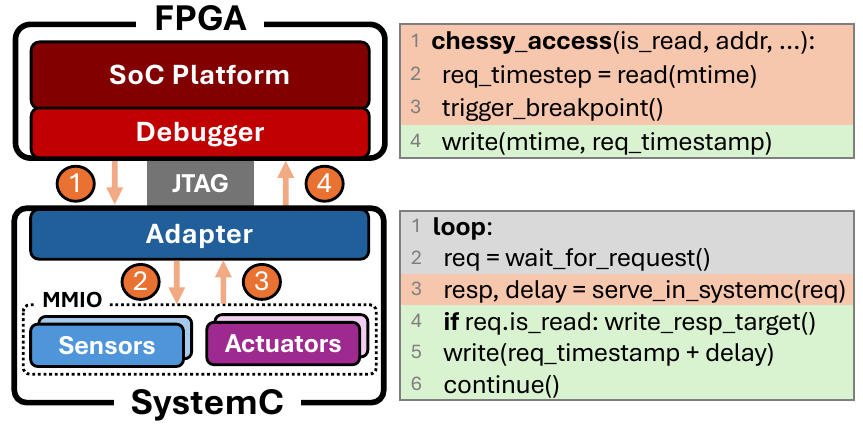}
    \vspace{-0.7cm}
    \caption{Overview of CHESSY. The background shades in the pseudocode (right) indicate the timestamp observed by the corresponding hardware (left): \textit{gray} = previous request, \textit{orange} = current request, \textit{green} = current request + simulated peripheral delay.}
    \label{fig:chessy}
    \vspace{-0.5cm}
\end{figure}

Figure~\ref{fig:chessy} illustrates CHESSY, which connects a SystemC VP %
to an FPGA-based target through a debugger-mediated link. %

The FPGA runs an unmodified design containing a processor and (a subset of) its fully-digital on-chip peripherals, while SystemC, running on the host, simulates the rest of the Memory-Mapped I/O (MMIO) peripherals, especially those interacting with the external world, such as sensors and actuators. A host-side adapter links the two environments by intercepting I/O requests at dedicated breakpoints and maintaining time synchronized through controlled updates to the FPGA’s timer registers (e.g., \texttt{mtime}).
This lets the software running on the FPGA perceive realistic peripheral latencies,  even though the latter are simulated on the host. 

Figure~\ref{fig:chessy} captures this process: \circled{1} whenever the FPGA wishes to communicate with a simulated device, it issues a \texttt{SystemCRequest} via a wrapper function \texttt{chessy\_access}; \circled{2} this request is intercepted by the host adapter, a SystemC module, that dispatches it to the appropriate peripheral; \circled{3} the peripheral SystemC model computes the response, after a specified \texttt{simulated\_delay}; finally, \circled{4} the adapter returns the result to the FPGA and updates timer registers accordingly. The only FPGA-side requirement to realize this behavior is a debugger server accessible through standard physical connections such as JTAG, providing basic breakpoint handling and memory/register access. This avoids RTL modifications, reduces hardware integration effort, and keeps the framework portable across FPGA vendors, processor models, debuggers, phy links, and simulation environments.

\vspace{-0.2cm}
\subsection{Communication and synchronization details}
CHESSY synchronizes FPGA and VP executions at each interaction (e.g., a read/write request from the FPGA-emulated core to a peripheral virtualized in SystemC). This minimizes synchronization overheads while maintaining timing accuracy equal to the resolution of the timer register present in the FPGA-emulated system (a requirement of our approach). Namely, the interaction works as follows. 

The \texttt{chessy\_access} function is implemented as a SW stub running on the FPGA, and it packages each communication request into a transaction structure containing operation type (\texttt{is\_read}), target address (\texttt{addr}), data pointer (\texttt{data\_ptr}), transfer size (\texttt{size\_bytes}), and local timestamp (\texttt{timestamp\_us}, read from \texttt{mtime}). When the request is ready, a breakpoint notifies the host. 

On the host, the adapter controls the debugger to monitor for breakpoints on \texttt{chessy\_access}, and reads the request transaction data structure; for write operations, it fetches the payload from the FPGA memory pointed to by \texttt{data\_ptr}; for read operations, it prepares to write the response back after processing.
Data transfer occurs over JTAG via the GDB \texttt{restore} (write) and \texttt{dump} (read) commands.

Next, the SystemC global simulation time is advanced until \texttt{timestamp\_us}.
The adapter then encapsulates the request into a high-level \texttt{SystemCRequest} and forwards it to the appropriate SystemC peripheral, determined by the target address. Peripheral models compute the functional response and advance the simulated time to model the peripheral's latency.

When a read reply is ready, the adapter writes the result into FPGA memory and advances time (\texttt{mtime}) to \(\texttt{timestamp\_us} + \texttt{simulated\_delay}\), ensuring that the FPGA’s perception of time reflects the simulated behavior.

\vspace{-0.2cm}
\section{Results}
\label{sec:results}

To validate the proposed approach, we emulated the RISC-V-based Astral platform \cite{astral} on a AMD VCU118 FPGA, connected to a Linux workstation running the MESSY \cite{messy} SystemC-VP for peripherals simulation. The workstation is equipped with a quad-core CPU, 32GB of RAM, an RV64 GNU GCC / GDB toolchain, and an RTL simulator (Siemens QuestaSim 2022.3) for baseline comparisons.

\begin{figure}
    \centering
    \includegraphics[width=\linewidth]{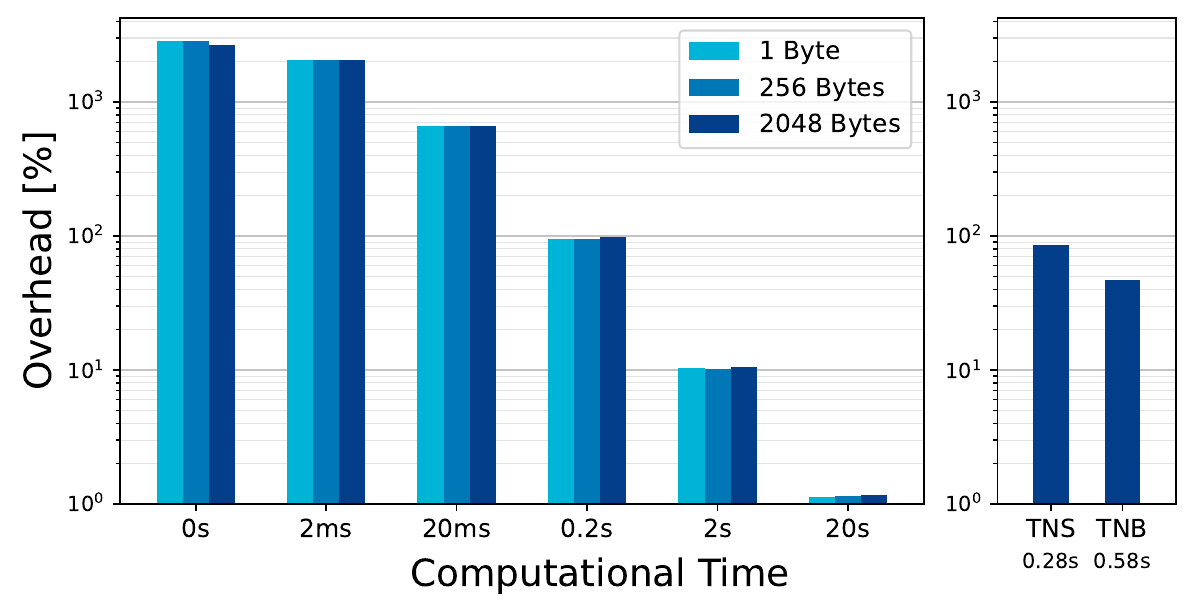}
    \vspace{-0.7cm}
    \caption{Simulation time overhead [\%] for different transfer sizes and interaction intervals. The two rightmost bars show overheads for 14\,M (TNS) and 28\,M (TNB) cycle variants of TempoNet (280\,ms and 560\,ms at 50\,MHz).}
    \vspace{-0.5cm}
    \label{fig:overhead-plot}
\end{figure}

The simulation time overhead introduced by CHESSY was evaluated against a pure FPGA setup (i.e., without VP communication), using a periodic read--compute--write benchmark with a configurable delay to emulate a generic CPS workload. The benchmark was run while sweeping the computation-to-I/O ratio by varying: \textbf{(i)} the data transfer size, from a few bytes up to several kilobytes, and \textbf{(ii)} the computation delay, from zero to millions of cycles, thereby controlling the access frequency to VP-simulated models. The results, reported on the left side of Fig.~\ref{fig:overhead-plot}, show that the overhead is nearly insensitive to transfer size, with relative variation always below \(4\%\), since data exchanges are implemented as host memory dump/restore operations. The total overhead clearly depends on the frequency of FPGA-VP interactions, but reduces to acceptable values for compute-intensive workloads, which are those that necessitate FPGA-acceleration in the first place.
On average, each access incurs a nearly constant overhead of less than \(100\,\text{ms}\).

We then assessed CHESSY's effectiveness on a realistic biomedical use case, including an Electromyography (EMG) sensor and a robotic arm, both modeled in SystemC, connected to the RISC-V SoC mapped on the FPGA. The application SW reads EMG data, performs gesture recognition via the TempoNet Neural Network (using either a small \(\approx 4\text{k}\)-parameter / 14\,M-cycles variant or a larger \(\approx 9\text{k}\)-parameter / 28\,M-cycles variant), and sends commands to the robotic arm \cite{temponet}. As with the synthetic benchmark, we measure the average overhead relative to an FPGA-only baseline, reported on the right side of Fig.~\ref{fig:overhead-plot}. The 14\,M-cycle network exhibits an overhead of about \(86\%\), which reduces to 47\% for the 28\,M-cycle one. Hence, for realistic applications with similar compute-to-access ratios, the total time remains below \(2\times\) that of pure FPGA execution. 

Larger and more compute-intensive workloads, common in modern embedded ML, benefit even more: the relative overhead drops to nearly $1\%$ for workloads longer than $20\,\text{s}$. For extremely access-intensive use cases, the protocol remains usable, although simulation may become up to $30\times$ slower. Overall, CHESSY adds modest performance overhead, especially for compute-bound applications, while retaining practical efficiency even in I/O-heavy scenarios.

Finally, to relate CHESSY's performance to a traditional cycle-accurate RTL simulation, the biomedical workload was also executed on QuestaSim. As expected, the FPGA-accelerated setup (including the HIL-based CHESSY protocol) achieves speedups greater than \(2500\times\) over RTL.

\vspace{-0.2cm}
\section{Conclusions}
\label{sec:conclusions}

We presented CHESSY, an open-source, FPGA-agnostic hybrid prototyping framework that does not require RTL modifications. Through a lightweight JTAG-based mechanism, it allows FPGA-emulated hardware to interact with SystemC-modeled peripherals while preserving realistic timing behavior. 
Experiments show that CHESSY provides over three orders of magnitude speedup compared to RTL simulation, with total execution time remaining within $2\times$ that of pure FPGA emulation for realistic applications.

\newpage
\bibliographystyle{IEEEtran}
\bibliography{references}

\end{document}